# X-ray ionization yields and energy spectra in liquid argon

A. Bondar,[a,b] A. Buzulutskov,[a,b,*] A. Dolgov,[b] L. Shekhtman,[a,b] A. Sokolov[a,b]

[a]*Budker Institute of Nuclear Physics SB RAS, Lavrentiev avenue 11, 630090 Novosibirsk, Russia*
[b]*Novosibirsk State University, Pirogov street 2, 630090 Novosibirsk, Russia*

*E-mail:* `A.F.Buzulutskov@inp.nsk.su`

ABSTRACT: The main purpose of this work is to provide reference data on X-ray ionization yields and energy spectra in liquid Ar to the studies in the field of Cryogenic Avalanche Detectors (CRADs) for rare-event and other experiments, based on liquid Ar detectors. We present the results of two related researches. First, the X-ray recombination coefficients in the energy range of 10-1000 keV and ionization yields at different electric fields, between 0.6 and 2.3 kV/cm, are determined in liquid Ar based on the results of a dedicated experiment. Second, the energy spectra of pulsed X-rays in liquid Ar in the energy range of 15-40 keV, obtained in given experiments including that with the two-phase CRAD, are interpreted and compared to those calculated using a computer program, to correctly determine the absorbed X-ray energy. The X-ray recombination coefficients and ionization yields have for the first time been presented for liquid Ar in systematic way.

KEYWORDS: X-ray recombination coefficients and ionization yields in liquid argon; X-ray energy spectra in liquid argon.

---

[*] Corresponding author.

# 1. Introduction

The main purpose of this work is to provide reference data on X-ray ionization yields and energy spectra in liquid Ar to the studies in the field of rare-event and other experiments, based on liquid Ar detectors. In particular, our group has a long-term activity in the field of two-phase Cryogenic Avalanche Detectors (CRADs) for dark matter search and coherent neutrino-nucleus scattering experiments [1] and their energy calibration [2],[3]. In the course of these developments, two X-ray sources were constantly used to produce ionization in liquid Ar: a 59.5 keV X-ray line of the $^{241}$Am radioactive source and the original pulsed X-ray tube with a continuous X-ray spectrum in the range of 15-40 keV (0,11BSV7-Mo, [4],[5]). For many years, these X-ray sources were successfully used to measure physical quantities characterizing CRAD performances: the charge gain [6], the spatial resolution [7], the scintillation light yield in the Near Infrared (NIR) [8] and the ionization yield of nuclear recoils [3].

On the other hand, the X-ray ionization yields and energy spectra in liquid Ar, referred to in our previous [3],[4],[7],[8] and current [9] works, have not been properly reflected in the literature. In this report we fill this gap, namely we present the results of two related researches in the field. First, the X-ray recombination coefficients in the energy range of 10-1000 keV and ionization yields at different electric fields are determined in liquid Ar, based on the results of a dedicated experiment. Second, the energy spectra of pulsed X-rays in liquid Ar measured in given experiments, in the energy range of 15-40 keV, are interpreted and compared to those calculated using a computer program [10], to correctly determine the absorbed X-ray energy.

It should be remarked that while the data on X-ray energy spectra are given here mostly for the reference purposes, the data on the recombination coefficients and ionization yields presented in the current work have a scientific novelty. Indeed, in contrast to Xe, little has been known so far about the X-ray ionization yield in liquid Ar at energies below 300 keV [2],[11]. Accordingly, these data might be of particular interest for the energy calibration of dark matter and neutrino detectors using liquid Ar detection medium.

# 2. Experimental procedures

Fig. 1 shows schematic views of two experimental setups to perform the two studies discussed above. The left panel presents a dedicated liquid Ar ionization chamber, to measure the ionization yield at lower energies using a pulsed X-ray tube. The right panel presents a two-phase CRAD in Ar with THGEM/GAPD-matrix charge/optical readout, to measure its amplitude and coordinate characteristics, where either a pulsed X-ray tube or a 59.5 keV X-ray line from the $^{241}$Am source was used to provide ionization in liquid Ar induced by X-rays of certain energies. The latter experimental setup was identical to that of ref. [7] where the first results on the performance of the combined multiplier composed of THGEMs [12] and a matrix of Geiger-mode APDs (GAPDs, [13]) in the two-phase CRAD were presented; the setup photograph is shown in Fig. 2.

Regarding cryogenic equipment and high voltage supply, the setups were similar to those described elsewhere [4],[6],[7],[8]. Each experimental setup consisted of a 9 l vacuum-insulated cryogenic chamber filled with liquid Ar. The electron life-time in the liquid was above 13 μs [6]. This corresponds for example to the electron mean free path of 29 mm at an electric field of 1.25 kV/cm [6], thus providing a rather uniform response across the liquid Ar layers used in this work. The measurements were conducted at 87 K either in liquid Ar, in the first setup (Fig. 1, left), or two-phase Ar, in the second setup (Fig. 1, right).

In both experimental setups, the pulsed X-rays were produced by a pulsed X-ray tube with Mo anode (0,11BSV7-Mo, [4],[5]). The tube was operated at a voltage of 40 kV and an anode current of 2.5 mA in a pulsed mode with a frequency of 240 Hz. The latter was provided by a gating grid, which was gated by a pulse generator. The pulse generator also provided a trigger for reading out the data. The cryogenic chamber was irradiated from outside practically uniformly across the active area, through a collimator and two aluminium windows at the chamber bottom, each 1 mm thick and of a diameter of 5 cm.

In the first experimental setup, the pulsed X-rays reached the liquid Ar ionization chamber after passing through a THGEM plate, 14 mm thick liquid Ar layer and another THGEM plate, acting as additional X-ray filters to the two Al input windows: see Fig. 1 (left). In the second experimental setup, the pulsed X-rays reached the cathode gap after passing through a THGEM-plate electrode laying on the chamber bottom and acting as a cathode. The THGEMs had the following geometrical parameters: 0.4 mm thick dielectric (FR4) clad by 35 μm thick copper on each side, hole diameter of 0.5 mm, hole rim (without copper cladding) of 0.1 mm and hole pitch of 0.9 mm.

The appropriate spectra of X-rays absorbed in the sensitive volume, namely in the ionization gap in the first setup and in the cathode gap in the second setup, were calculated using a computer program XOP [10]; they are shown in Fig. 3. One can see that the average energy of absorbed X-rays amounts to 35 keV in the first setup and 25 keV in the second setup. In section 4 the latter spectrum will be compared to that obtained in experiment.

In the first experiment setup (Fig. 1, left), the X-ray pulse had a sufficient power to provide measurable ionization charge, normally having values of tens of thousands electrons in liquid Ar. In addition, it was sufficiently fast, having a width of 0.5 μs. The ionization signal was recorded in the ionization chamber (parallel-plate gap) with an active area of 30×30 mm$^2$ and thickness of 2 mm. The gap was formed by a THGEM-plate electrode and a wire grid as shown in Fig. 1 (left). The charge (ionization) signal was read out from the anode electrode using a charge-sensitive preamplifier followed by research amplifier with an overall shaping time of 10 μs, with the electronic noise corresponding to the Equivalent Noise Charge of 4000 e. The amplifiers were placed outside the cryogenic chamber. The signals were digitized and recordered for further off-line analysis with a TDS5032B digital oscilloscope, by the trigger provided by the X-ray tube pulse generator.

In the second experimental setup (Fig. 1, right and Fig. 2), the cryogenic chamber included a cathode electrode, immersed in a 8 mm thick liquid Ar layer, and a double-THGEM multiplier with an active area of 10×10 cm$^2$, placed in the gas phase above the liquid and optically read out by a 3×3 matrix of GAPDs in the NIR. The cathode and the THGEMs were biased through a resistive high-voltage divider, placed outside the cryostat. In all the measurements, the electric field within liquid Ar was kept at 1.75 kV/cm. In case of the pulsed X-ray irradiation, a steel cylindrical collimator with a hole diameter of 2 mm and thickness of 10 mm was used. In addition to the pulsed X-rays, the detector could be irradiated from outside by X-rays from a $^{241}$Am source providing a 59.5 keV line at a rate of a few Hz. The charge signal was recorded from the last electrode of the second THGEM using a charge-sensitive preamplifier followed by research amplifier with an overall shaping time of 10 μs (similar to the first setup). The DAQ system included an 8-channel Flash ADC CAEN V1720 (12 bits, 250 MHz): the optical signals from 7 GAPDs (after the fast amplifiers) and the charge signal from the double-THGEM multiplier were digitized and stored in a computer for further off-line analysis.

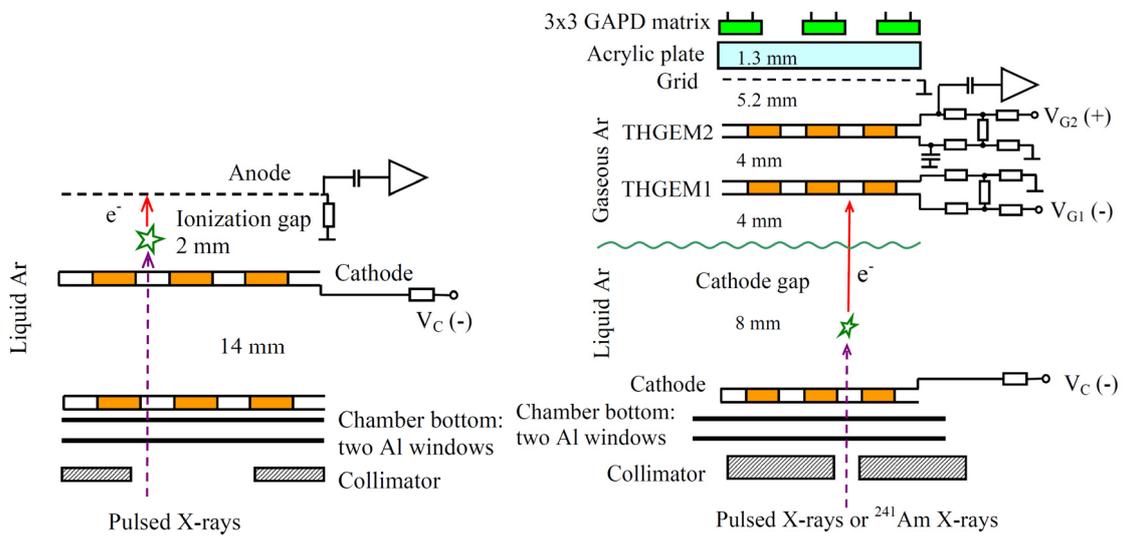

Fig.1 Left: schematic view of a dedicated liquid Ar ionization chamber to measure the X-ray ionization yield. Right: schematic view of a two-phase CRAD in Ar with THGEM/GAPD-matrix charge/optical readout to study its amplitude and coordinate characteristics [7] .

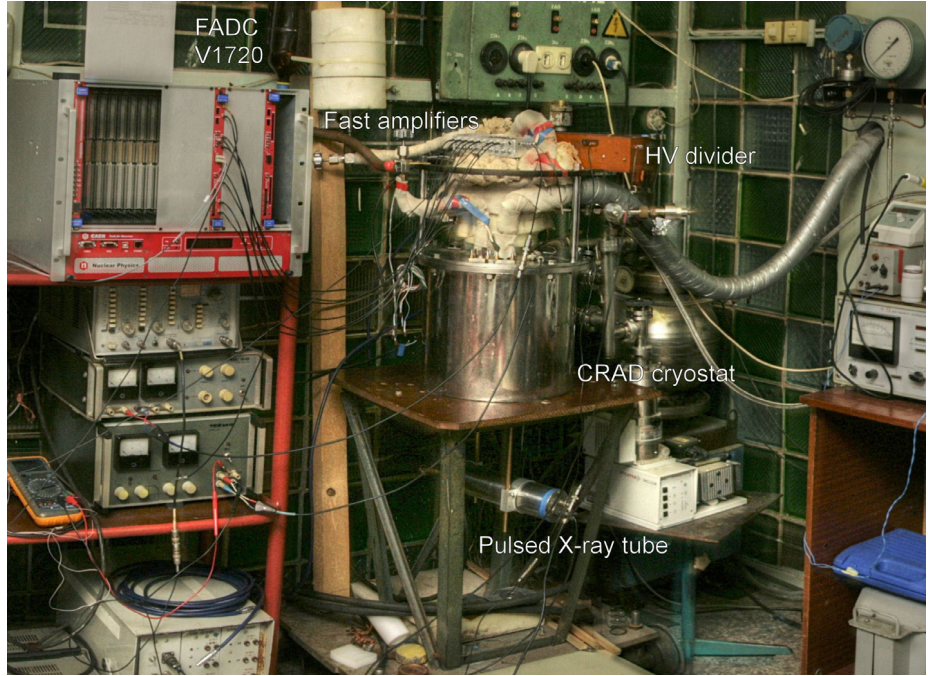

Fig.2 Photograph of the two-phase CRAD in Ar with THGEM/GAPD-matrix charge/optical readout of Fig. 1 (right) [7]. The CRAD cryostat, pulsed X-ray tube, HV divider box, fast amplifiers box and flash ADC module V1720 in the VME crate of the DAQ system are indicated in the figure.

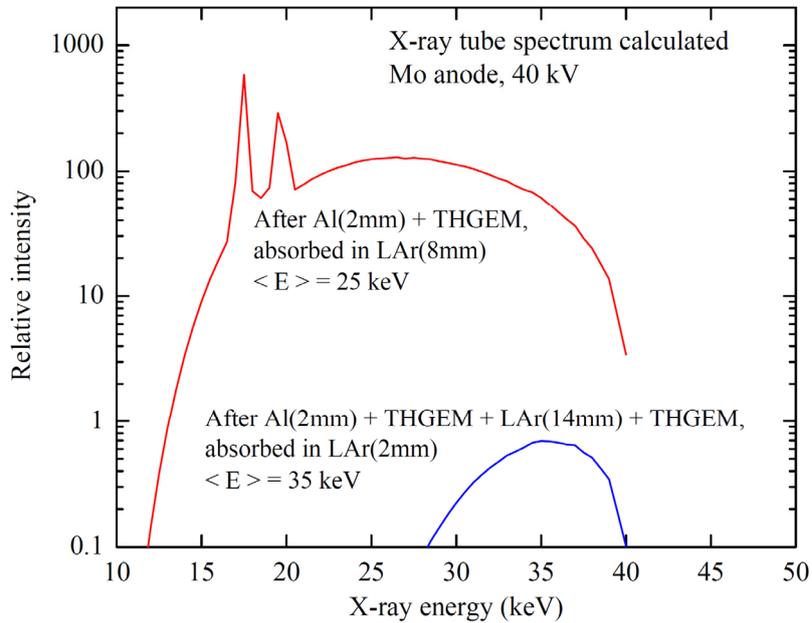

Fig.3 Calculated spectra of pulsed X-rays absorbed either in the ionization gap of the liquid Ar ionization chamber or in the cathode gap of the two-phase CRAD of Fig. 1 (left) and (right) respectively, i.e. after filtering with 2 mm thick Al windows and with either combination of THGEM + 14 mm of LAr + THGEM or that of just THGEM.

## 3. Ionization yields

The ionization yield from a track of a particle absorbed in the noble-gas liquid ($Q_y$) is defined as follows:

$$Q_y = n_e / E_0 \ . \tag{1}$$

Here $n_e$ is the number of electrons escaping recombination with positive ions; it depends on the nature of the particle collision with atoms (electron or nuclear recoils), on the energy deposited by a particle in the liquid ($E_0$) and on the electric field in the liquid ($\mathcal{E}$). $n_e$ is always smaller than the initial number of ion pairs produced in the liquid by a particle ($N_i$). In the absence of a complete recombination model, it is generally accepted that the following parametrization works well [2],[14],[15]:

$$n_e = N_i / (1 + k/\mathcal{E}) \ , \tag{2}$$

where $k$ is a fitting constant, often called recombination coefficient. These equations are valid for both electron recoils, induced by electron or gamma-ray irradiation, and nuclear recoils.

In this section, we determine the recombination coefficients and ionization yields induced by X-ray absorption in liquid Ar for the energies and electric fields relevant to our previous [3],[6],[7] and current [9] studies, in particular in the energy range of 10-1000 keV. The data on the X-ray ionization yields in liquid Ar at an electric field of 2.3 kV/cm have been already used in ref. [3], to calibrate the ionization charge scale when measuring the ionization yield of nuclear recoils in liquid Ar. In addition, the similar data will be used in our forthcoming publications on the performance of the two-phase CRAD with THGEM/GAPD-matrix multiplier and that with electroluminescence gap [9], at a field of 1.75 kV/cm and 0.6 kV/cm respectively. Accordingly, in this section we determine the X-ray ionization yields in liquid Ar for the electric field values of 0.6, 1.75 and 2.3 kV/cm, using the energy dependence of the ionization yields in the energy range of 10-1000 keV. Note that the experimental results of this section are obtained with the detector of Fig. 1 (left).

To determine this energy dependence one should know in turn the recombination coefficients at several energy points. While for liquid Xe there are enough data on the recombination coefficients in the energy range of interest [2],[14],[16],[17],[18] and on those of ionization yield [19],[20], little was known about those in liquid Ar [11],[17],[21]. This is seen from Table 1 showing the recombination coefficients in liquid Ar and Xe for X-ray- or electron-induced ionization. This is also seen from Fig. 4 showing the relative ionization yields in liquid Ar and Xe, calculated from Eq. (2) using those recombination coefficients, as a function of the X-ray or electron energy, at an electric field of 1.75 kV/cm. Here all the Xe recombination coefficients and that of Ar at 976 keV are reproduced from Table 2.6 of ref. [14]. The recombination coefficient for Ar at 364 keV was obtained ourselves from the data of ref. [21]. One can see that the ionization yield decreases monotonically with decreasing energy.

| Liquid | X-ray or electron energy $E_0$ (keV) | Recombination coefficient $k$ (V/cm) | Reference |
|---|---|---|---|
| Ar | 35 | 1820±110 | This work |
| Ar | 364 | 584±40 | Deduced from [21] |
| Ar | 976 | 560±10 | [17] |
| Xe | 15.3 | 2370±150 | [16] |
| Xe | 17.3 | 2100±300 | [16] |
| Xe | 21.4 | 1800±80 | [16] |
| Xe | 550 | 410±30 | [17] |
| Xe | 662 | 420±30 | [18] |

Table 1. Recombination coefficients in liquid Ar and Xe for X-ray- or electron-induced ionization in the energy range of 10-1000 keV. The Xe recombination coefficients and that of Ar at 976 keV are reproduced from Table 2.6 of ref. [14].

As seen from Table 1, there are only two data points in energy for Ar that can be found in the literature. The third data point is obtained in the current work, in the lower energy domain, from the measurements in the liquid Ar ionization chamber of Fig 1 (left). The result of this measurement is shown in Fig. 4 showing the anode signal charge per X-ray pulse, as a function of the electric field, produced by pulsed X-rays with the average energy of 35 keV (their calculated energy spectrum is shown in Fig. 3). The measurement errors shown in the figure were due to X-ray photon statistics (amounting to about 100 photons per pulse multiplied by typically 16 pulses taken to measure one data point) and those of electronics noise. The curve in the figure is the recombination model fit of Eq. (2) to the data points, resulting in the following value of the recombination coefficient: $k$=1820±110 V/cm, at the X-ray energy of 35±3 keV.

In liquid Xe, where there are enough experimental data in the energy range of interest, the relative ionization yield dependence on energy is perfectly described by a function

$$n_e / N_i = a / (1+b/E_0) \quad , \tag{3}$$

with two parameters ($a$ and $b$): see Fig. 4. Accordingly, in liquid Ar we used the similar function, firstly, to describe the experimental data and, secondly, to interpolate the data to the energy point of interest.

The results are presented in Figs. 4 and 6 showing the relative ($n_e/N_i$) and absolute ($Q_y$) ionization yields in liquid Ar as a function of energy. To obtain the latter, we used a *W*-value (energy needed to produce one ion pair) and its definition,

$$W = E_0 / N_i \quad ; \tag{4}$$

*W*=23.6 eV for electron recoils in liquid Ar (see table 2 in ref. [2]).

| Electric field (kV/cm) | Energy dependence of the relative ionization yield |
|---|---|
| 0.6 | $n_e / N_i = 0.540\pm0.007 / (1 + 41\pm5/E_0[\text{keV}])$ |
| 1.75 | $n_e / N_i = 0.774\pm0.005 / (1 + 20\pm2/E_0[\text{keV}])$ |
| 2.3 | $n_e / N_i = 0.818\pm0.004 / (1 + 16\pm2/E_0[\text{keV}])$ |

Table 2. Relative ionization yield dependence on energy in liquid Ar for electron recoils due to X-ray irradiation, at different electric fields used in our experiments.

| Electric field (kV/cm) | X-ray energy $E_0$ (keV) | Number of ionization electrons escaping recombination $n_e$ (e$^-$) | Ionization yield $Q_y$ (e$^-$/keV) |
|---|---|---|---|
| 0.6 | 59.5 | 810±40 | 13.6 |
| 1.75 | 25 | 460±20 | 18.4 |
| 1.75 | 59.5 | 1470±40 | 24.7 |
| 2.3 | 59.5 | 1630±50 | 27.4 |

Table 3. Predicted number of ionization electrons escaping recombination and absolute ionization yield in liquid Ar for electron recoils due to X-ray irradiation, at different energies and electric fields used in our experiments.

The relative ionization yield dependence on energy and the predicted number of ionization electrons escaping recombination in liquid Ar, as well as the predicted absolute ionization yield, at different electric fields and X-ray energies used in our experiments, are summarized in Tables 2 and 3. The values in Table 3 are obtained by interpolation using the appropriate energy dependence of Table 2. For example, for 59.5 keV X-rays the number of detected ionization electrons in liquid Ar at a field of 1.75 kV/cm is predicted to be $n_e$=1470±40 e$^-$.

It should be remarked that we deliberately ignored here the low-energy region, below 10 keV, since the energy dependence of the ionization yield below this energy might change its behaviour. In particular in liquid Xe, at further energy decrease a minimum is predicted first, and then the growth of the yield [19],[20]. Moreover, such a behaviour is definitely observed in liquid Ar when the higher energy data are combined with those of the lower energy: see Fig. 7 where the data on the absolute ionization yield in liquid Ar are summarized for electron recoils, at an electric field of 2.4 kV/cm.

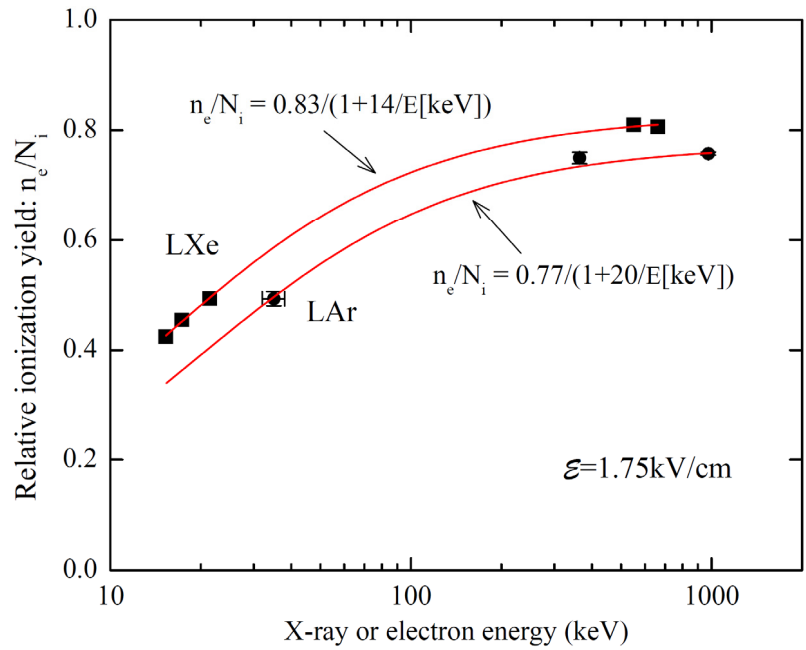

Fig.4 Relative ionization yields in liquid Ar and Xe for electron recoils, due to X-ray or electron irradiation, as a function of the X-ray or electron energy, at an electric field of 1.75 kV/cm.

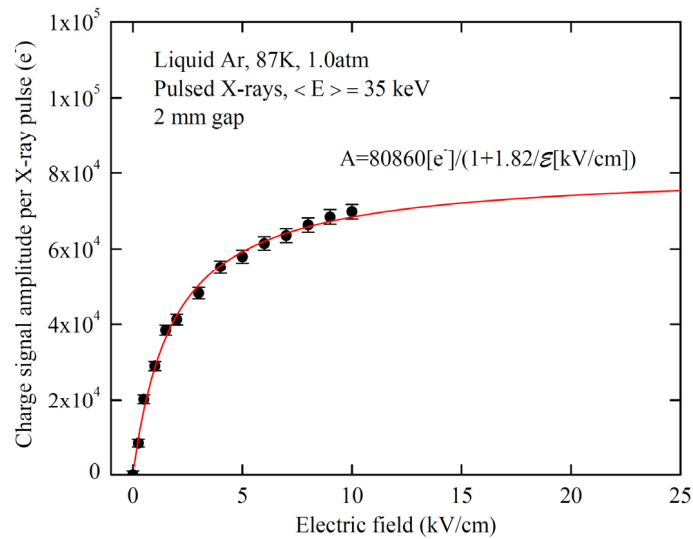

Fig.5 Anode signal charge expressed in electrons, per X-ray pulse, as a function of the electric field in the liquid Ar ionization chamber of Fig. 1 (left), produced by pulsed X-rays with the average energy of 35 keV. The curve is the recombination model fit to the data points.

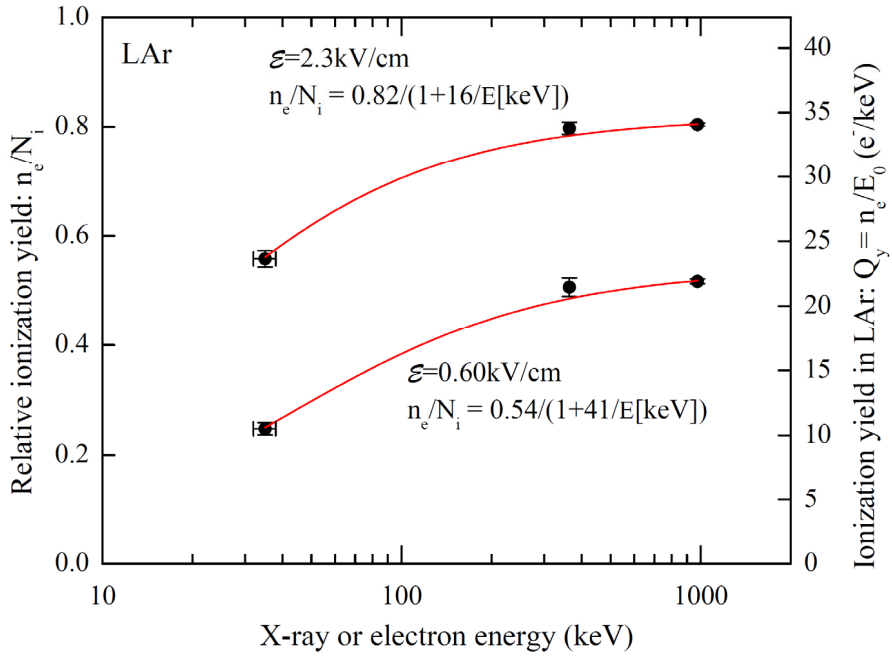

Fig.6 Relative (left scale) and absolute (right scale) ionization yields in liquid Ar for electron recoils, due to X-ray or electron irradiation, as a function of the X-ray or electron energy, at electric fields of 0.6 and 2.3 kV/cm.

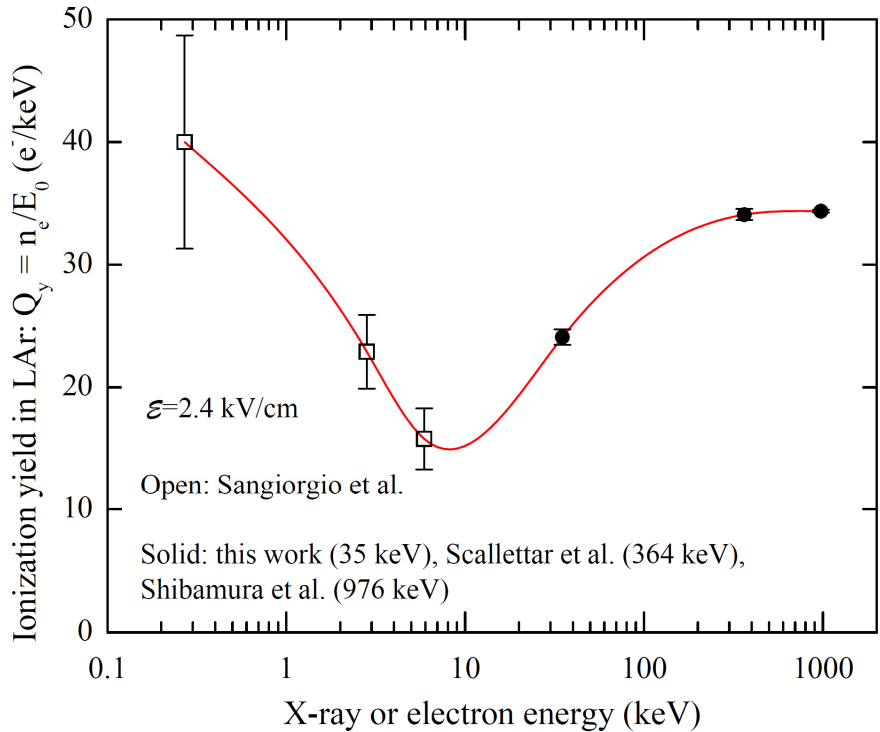

Fig.7 Summary of absolute ionization yields in liquid Ar for electron recoils, due to X-ray or electron irradiation, as a function of the X-ray or electron energy, at an electric field of 2.4 kV/cm. The data points were directly measured at lower energies in Sangiorgio et al. [11] (at 0.27 keV, 2.8 keV and 5.9 keV) and calculated using recombination coefficients at higher energies in this work (at 35 keV), Scallettar et al. [21] (at 364 keV) and Shibamura et al. [17] (at 976 keV). The solid line is the spline function fit to the combined data points.

## 4. Energy spectra

Fig. 3 shows two calculated energy spectra of pulsed X-rays used in the first and second experimental setup; the first spectrum has been already employed in section 3. The second spectrum will be employed in this section.

In the second experimental setup, the multi-channel optical readout of the two-phase CRAD in Ar, using a combined THGEM/GAPD-matrix multiplier, was for the first time demonstrated; the appropriate results were presented in a short paper [7]. The detector had a spatial resolution of 2.5 mm (FWHM), which is rather good for two-phase detectors, and a low detection threshold. In particular, at a rather moderate charge gain of 160, the yield of the combined THGEM/GAPD-matrix multiplier was claimed to be equal to ~80 photoelectrons per 20 keV X-ray absorbed in liquid Ar. In fact, the X-ray energy spectra in liquid Ar were not even presented there [7]. In this section we fill this gap, namely we compare the measured pulsed X-ray energy spectrum to that calculated in Fig. 3 and thus specify more precisely the energy of the absorbed X-ray photons. These data are essential for the accurate determination of both the combined multiplier yield and detection threshold of the two-phase CRAD; these will be used in our forthcoming publication where more elaborated results will be presented.

In the case of pulsed X-rays, the X-ray tube was moved away, reaching a relatively large distance (of ~50 cm) from the collimator, to provide an operation in X-ray photon counting mode, with a rather small deposited energy per pulse, of the order of 20 keV. The pulsed X-ray energy spectrum measured in these conditions is shown in Fig. 8, along with the spectrum from the X-rays of the $^{241}$Am source. The latter was used to calibrate the energy scale and to determine the energy resolution, amounting to about $\sigma/E$=30%, using its 59.5 keV X-ray line. Note that the average energy of the pulsed X-ray spectrum measured using this calibration is equal to 19 keV for the whole spectrum, including the high-energy tail.

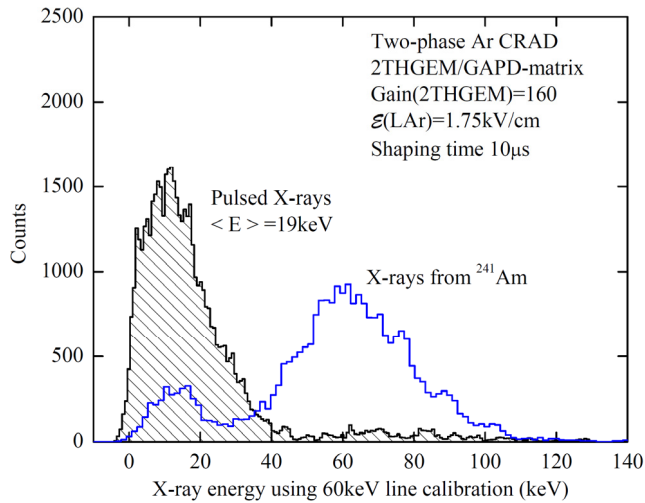

Fig.8 Energy distributions of charge signals from the double-THGEM in the two-phase CRAD in Ar with 2THGEM/GAPD-matrix multiplier of Fig. 1 (right), at a double-THGEM gain of 160. The signals were induced by X-rays from the $^{241}$Am source and pulsed X-ray tube. The energy scale is calibrated using a 60 keV line of the $^{241}$Am source. The electric field within liquid Ar was 1.75 kV/cm.

In fact, the real energies of the absorbed X-ray photons differ from those of "measured". Indeed, the "measured" pulsed X-ray energy, using the calibration at different energy point, is not equal to its real energy due to the dependence of the X-ray ionization yield on the energy: see Fig. 4 and Table 3. To correctly determine the real X-ray energy, we should compare the measured spectrum to the simulated one, in the frame of a certain computation model.

Namely, we first took the calculated spectrum of Fig. 3 (after filtering with 2 mm thick Al windows and the THGEM and absorbed in 8 mm thick liquid Ar layer) and reduced its energy, to correct it for the ionization yield decrease in comparison with that of the energy calibration point (at 59.5 keV). This correction was introduced by the equation $n_e/N_i=0.78/(1+20/E_0[keV])$ of Table 2, for the electric field of 1.75 kV/cm, resulting in some intermediate version of the spectrum for the single-photon events.

The next step was to play Poisson statistics for X-ray photon counting, in order to take into account the zero-photon and multiple-photon events. This step included a Monte-Carlo simulation of 1 photon, 2 photon and 3 photon event spectra, using the intermediate spectrum for 1 photon events obtained before, and their further convolution with the Gauss distribution, to correct for the energy resolution contribution ($\sigma/E=30\%$). For the 0 photon (background) contribution we took a Gaussian distribution, its parameters being obtained from spectrum adjusting to the left edge of the experimental spectrum. Then these spectra were summarized, their amplitudes being linked according to Poisson statistics. Finally, this computation model spectrum was fitted to that of experimental, using the Poisson statistics mean (equal to the average photon number $<N_{ph}>$) and an overall scaling factor as the only free parameters.

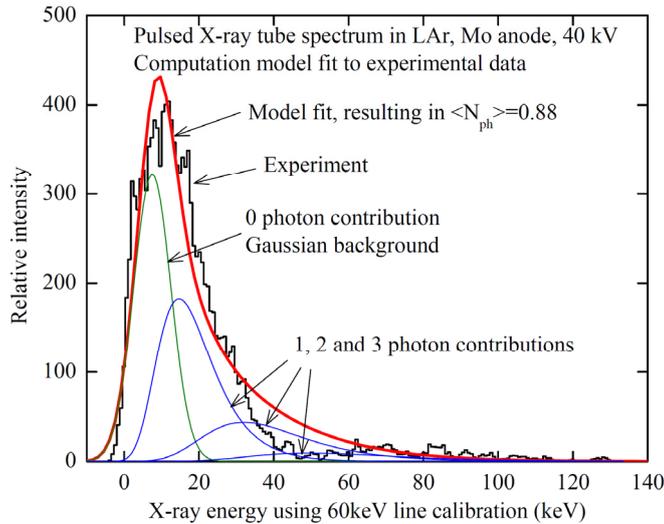

Fig.9 Energy distributions of charge signals in the two-phase CRAD in Ar with 2THGEM/GAPD-matrix multiplier of Fig. 1 (right) induced by pulsed X-rays. The measurement conditions are those of Fig. 8. The energy scale is calibrated using a 60 keV line of the $^{241}$Am source. The experimental spectrum is compared to that of computation model, the latter being computed for X-ray absorption in 8 mm liquid Ar layer after filtering with 2 mm thick Al and THGEM. The computed spectrum was first corrected for the energy dependence of the ionization yield, the energy resolution ($\sigma/E=30\%$) and Poisson statistics for X-ray photon counting, and then fitted to the experimental spectrum, the average photon number (<Nph>) being a free parameter. The contributions of 0 photon, 1 photon, 2 photon and 3 photon events to the computed spectrum are indicated in the figure.

The fit resulted in $<N_{ph}>$=0.88, i.e. we had roughly one X-ray photon per pulse. The computed spectrum is shown in Fig. 9 in comparison with that of experimental. The contributions of the 0 photon, 1 photon, 2 photon and 3 photon event spectra to the computation model spectrum are also presented in the figure. One can see that the computed spectrum generally reflects the shape and the position of the measured spectrum, confirming simulated characteristics of the absorbed X-ray radiation, with its average photon energy of 25 keV.

## 5. Conclusions

In this work we present the results of two related studies. In the first study, we determined the X-ray recombination coefficients and ionization yields in liquid Ar, for the energies and electric fields relevant to our previous [3],[7] and present [9] studies in the field of two-phase CRADs for rare-event experiments, in particular in the energy range of 10-1000 keV and electric fields of 0.6, 1.75 and 2.3 kV/cm.

In the second study, the energy spectra of pulsed X-rays in liquid Ar in the energy range of 15-40 keV, obtained in given experiments including that with the two-phase CRAD, are interpreted and compared to those calculated using a computer program, to correctly determine the absorbed X-ray energy.

It should be remarked that the X-ray recombination coefficients and ionization yields have for the first time been presented here for liquid Ar in systematic way. These data might be of particular interest for the energy calibration of dark matter and neutrino detectors using liquid Ar detection medium.

## 6. Acknowledgements

This work consisted of two independent studies conducted on different experimental setups. The first study was supported in part by Russian Science Foundation (project N 14-50-00080). The second study was supported in part by the grant of Russian Foundation for Basic Research (15-02-01821).

## References

[1] A. Buzulutskov, JINST 7 (2012) C02025.

[2] V. Chepel, H. Araujo, JINST 8 (2013) R04001.

[3] A. Bondar et al., Europhys. Lett. 108 (2014) 12001.

[4] A. Bondar et al., JINST 7 (2012) P06015.

[5] Svetlana-X-ray company, http://svetlana-x-ray.ru.

[6] A. Bondar et al., JINST 8 (2013) P02008, and references therein.

[7] A. Bondar et al., Nucl. Instr. and Meth. A 732 (2013) 213 [arXiv:1303.4817].

[8] A. Bondar et al., JINST 7 (2012) P06014.

[9] A. Bondar et al., Europhys. Lett. 112 (2015) 19001.

[10] www.esrf.eu/computing/scientific/xop2.1/ or https://www1.aps.anl.gov/Science/Scientific-Software/XOP.

[11] S. Sangiorgio et al., Nucl. Instr. and Meth. A 728 (2013) 69.

[12] A. Breskin et al., Nucl. Instr. and Meth. A 598 (2009) 107, and references therein.

[13] D. Renker, E. Lorenz, JINST 4 (2009) P04004, and references therein.

[14] A.S. Barabash, A.I. Bolozdynya, Liquid ionization detectors, Energoatomizdat, Moscow, 1993 (in Russian).

[15] E. Aprile, A. Bolotnikov, A. Bolozdynya, T. Doke, Noble gas detectors, WIILEY-VCH, Weinheim 2006, and references therein.

[16] T.Ya. Voronova et al., Sov. Phys. Tech. Phys. 34 (1989) 825 [Zh. Tekh. Fiz. 59 (1989) 186].

[17] E. Shibamura et al., Nucl. Instr. and Meth. 131 (1975) 249.

[18] I.M. Obodovskii, S.G. Pokachalov, Sov. J. Low Temp. Phys. 5 (1979) 393.

[19] D.Yu. Akimov et al., JINST 9 (2014) P11014.

[20] M. Szydagis et al., JINST 8 (2013) C10003.

[21] R.T. Scallettar et al., Phys. Rev. A 25 (1982) 2419.